\documentclass[preprint,showpacs,preprintnumbers,amsmath,amssymb,showkeys]{revtex4}

\usepackage{graphicx}
\usepackage{dcolumn}
\usepackage{bm}

\begin{document}

\title{Discrete flavor symmetry and minimal seesaw mechanism}


\author{N. W. Park} \email{nwpark@jnu.ac.kr}
\affiliation{Department of Physics, Chonnam National University, Gwangju 500-757, Korea}
\author{K. H. Nam} \email{khnam@phys.cau.ac.kr}
\author{Kim Siyeon} \email{siyeon@cau.ac.kr}
\affiliation{Department of Physics,
        Chung-Ang University, Seoul 156-756, Korea}

\date{January 21, 2011}
\begin{abstract}
This work proposes a neutrino mass model that is derived using the minimal seesaw mechanism which contains only two right-handed neutrinos, under the non-abelian discrete flavor symmetry $\mathbb{S}_4\otimes\mathbb{Z}_2$. Two standard model doublets, $L_\mu$ and $L_\tau$, are assigned simultaneously to a $\mathbf{2}$ representation of $\mathbb{S}_4$. When the scalar fields introduced in this model, addition to the Standard Model Higgs, and the leptons are coupled within the symmetry, the seesaw mechanism results in the tri-bi-maximal neutrino mixing. This study examined the possible deviations from TBM mixing related to the experimental data.
\end{abstract}


\maketitle \thispagestyle{empty}


\section{Introduction}

The data of neutrino oscillation experiments revealed a very noticeable form of a mixing matrix with $0.50\leq|U_{e2}|\leq0.61, ~0.63\leq|U_{\mu3}|\leq0.79,$ and $|U_{e3}|\leq0.16,$ at the $90\%$ confidence level(CL). The current data also include the mass squared differences which are accompanied by solar and atmospheric neutrino oscillations, $\Delta m_{sol}^2 \simeq (7^{+10}_{-2})\times 10^{-5}eV^2$ and $\Delta m_{atm}^2 \simeq (2.5^{+1.4}_{-0.9})\times 10^{-3} eV^2$, respectively\cite{Nakamura:2010zzi}\cite{GonzalezGarcia:2007ib}. The central values of the elements of the Pontecorvo-Maki-Nakagawa-Sakata(PMNS) matrix are pointing a unique form of the matrix, which consists of $\sin\theta_{sol}=1/\sqrt{3}, ~ \sin\theta_{atm}=1/\sqrt{2},$ and $\sin\theta_{reactor}=0$. The three-flavor neutrino mixing matrix specified by those angle sizes is called tri-bi-maximal(TBM) mixing\cite{Harrison:2002er}, based on the idea that the neutrino mass matrix in a lepton flavor basis is an $\mathbb{S}_3$ group matrix and a canonical subgroup $\mathbb{S}_2$ plays a role of a $\mu-\tau$ interchange\cite{Harrison:2003aw}. Therefore, its elegant form suggests that the mass model has originated from a flavor symmetry. Models with various types of discrete flavor symmetries have been constructed
\cite{Zee:2005ut}\cite{Lam:2008rs}\cite{Zhang:2006fv}\cite{Bazzocchi:2009pv}.

A previous study considered a minimal seesaw model with only two righthanded neutrinos, where texture zeros and equalities in the Dirac mass matrix constrain the number of parameters and all the elements of the $2\times3$ Dirac matrix can be expressed in terms of low energy physical parameters\cite{Chang:2004wy}. In this study, $\mathbb{S}_4\otimes\mathbb{Z}_2$ symmetry is adopted to construct a minimal seesaw model, which is established with a small content of additional fields beyond the Standard Model(SM). The Higgs contents are extended to include three $SU(2)$ doublet scalars in one- or two- dimensional representations of $\mathbb{S}_4$ addition to the Standard Model Higgs, and then the model has a relatively simple Higgs potential, while the models with three righthanded neutrinos constructed in a group framework of the same kind should have included a rather complicated potential\cite{Zhang:2006fv}\cite{Bazzocchi:2009pv}. The model naturally represents $\mu-\tau$ symmetry and the properties of group $\mathbb{S}_4\otimes\mathbb{Z}_2$ lead the basis to the one in which the charged lepton mass matrix is diagonal. Since the exact $\mathbb{S}_4\otimes\mathbb{Z}_2$ symmetry for a given field content and given charge assignments is necessary, a meaningful issue can be the deviation from TBM mixing as a consequence of the symmetry breaking by a few vevs.

The outline is as follows. Section II introduces the framework within the flavor symmetry and the Lagrangian. The aspects that may arise during the symmetry breaking processes and their low energy effects since the seesaw mechanism are reported in Section III. Some extended effects of symmetry breaking, which are plausible for the deviation from TBM, as well as the experimental accessibility of the model are considered in Section IV. In the conclusion, the theoretical remarks on the model and an analysis on aspects of deviation from TBM mixing are summarized. The appendix contains the potential of all scalar particles along with their vacuum expectation values.

\section{Discrete flavor symmetry $\mathbb{S}_4\otimes\mathbb{Z}_2$}

The 5 irreducible representations of $\mathbb{S}_4$ are
    \begin{eqnarray}
    && \mathbf{1},~ \mathbf{1'},~ \mathbf{2},~ \mathbf{3},~ \mathbf{3'}.
    \end{eqnarray}
While the tensor products of $\mathbf{1}$, one of the 1-dim representations, are so trivial that they are simply $\mathbf{1}\times\mathbf{r} =\mathbf{r}\times\mathbf{1} = \mathbf{r},$ the tensor products of $\mathbf{1'}$, the other 1-dim representation, are as follows:
    \begin{eqnarray}
    && \mathbf{1'}\times\mathbf{1'} = \mathbf{1}~:~ ab, \label{prod1}\\
    && \mathbf{1'}\times\mathbf{2} = \mathbf{2}~:~
        \left(\begin{array}{r}ab_2\\-ab_1 \end{array}\right).
    \end{eqnarray}
In general, the elements of $\mathbf{2}$ are denoted by $a_i$ or $b_i$, for $i$ and $j=$ 1 to 2. According to the group theory details for the $\mathbb{S}_4$-II version in Ref.\cite{Merlo:2010mw}, the product of $\mathbf{2}$ with another $\mathbf{2}$ is
    \begin{eqnarray}
    && \mathbf{2}\times\mathbf{2} =
        \mathbf{1}+\mathbf{1'}+\mathbf{2}, \label{prod3} \\
    && \begin{array}{lcc}
        \mathbf{1} & : & (a_1b_1+a_2b_2) \\
        \mathbf{1'} & : & (a_1b_2-a_2b_1) \\
        \mathbf{2} & : &
        \left(\begin{array}{l}a_2b_2-a_1b_1\\a_1b_2+a_2b_1 \end{array}\right),
        \end{array} \label{prod5}
    \end{eqnarray}

The SU(2) lepton doublets $L_\alpha\equiv(L_e, L_\mu, L_\tau)$, the righthanded charged lepton singlets $E_R\equiv(e_R, \mu_R, \tau_R)$, and the Higgs scalar doublet $H$ in the Standard Model are assigned into such representations of $\mathbb{S}_4\otimes\mathbb{Z}_2$ as follows:
    \begin{eqnarray}
    \begin{array}{ccccc}
        \mathrm{Rep.} &\vline& (\mathbf{1},1)_F &(\mathbf{2},1)_F &(\mathbf{1},1)_F \\
    \hline
        (\mathbf{2},-1/2)_G &\vline& L_e &(L_\mu, L_\tau) & H \\ \vspace{2pt}
        (\mathbf{1},-1)_G &\vline& e_R & (\mu_R, \tau_R) & \\
    \hline
    \end{array} \label{charge_lepton}
    \end{eqnarray}
In the above table, the SU(2) representation and hypercharge of a field are denoted by the subscription `$G$' of the gauge symmetry, and $\mathbb{S}_4$ representation and  $\mathbb{Z}_2$ charge of the field are denoted by the subscription `$F$'. Under flavor symmetry, the charge assignment in Eq.(\ref{charge_lepton}) does not distinguish the $\mu$ flavor from the $\tau$ flavor, because $L_\mu$ and $L_\tau$ as well as $\mu_R$ and $\tau_R$ are in doublets. Therefore, this $\mathbb{S}_4\otimes\mathbb{Z}_2$ symmetry appears as $\mu-\tau$ symmetry.

To explain the light neutrino masses, additional Higgs scalars, $\Phi \equiv (\phi_1,\phi_2)$ and $\chi$, and right-handed neutrinos, $N_R \equiv (N_1,N_2)$, are introduced. Their representations under gauge symmetry and flavor symmetry are given in the following table.
    \begin{eqnarray}
    \begin{array}{cccccc}
        Rep. &\vline& (\mathbf{r}_1,q_1)_F &(\mathbf{r}_2,q_2)_F &(\mathbf{2},-1)_F &(\mathbf{1},-1)_F \\
    \hline
        (\mathbf{2},1/2)_{G} &\vline& & & (\phi_1, \phi_2) & \chi \\
        (\mathbf{1},0)_{G} &\vline& N_1 & N_2 & & \\
    \hline
    \end{array}\label{charge_nh}
    \end{eqnarray}
The flavor charges of $N_1$ and $N_2$, $(\mathbf{r}_1,q_1)_F$ and $(\mathbf{r}_2,q_2)_F$ will be assigned according to the mass type, normal hierarchy or inverted hierarchy. In general,
the Lagrangian constructed with the leptons and Higgs scalars in Eq.(\ref{charge_lepton}) and Eq.(\ref{charge_nh}) can be written as
    \begin{eqnarray}
    -\mathcal{L}=f_{ij}\overline{E}_R^iHL_\alpha^j
    +g_{jk}L_\alpha^j\Sigma\overline{N}_R^k +\frac{1}{2}M_{kl}\overline{N}_R^kN_R^l,
    \label{lagrangian}
    \end{eqnarray}
where $\Sigma=\{H, \Phi, \chi\}$. Each $i$ and $j$ runs 1 to 3, whereas each $k$ and $l$ runs 1 to 2.

The $\mathbb{S}_4\otimes\mathbb{Z}_2$ invariant Higgs potential is
    \begin{eqnarray}\begin{array}{cll}
    V   & = & V_H + m_\varphi^2 \Phi^\dagger\Phi + m_\chi^2 \chi^\dagger\chi  \\
        & + & \frac{1}{2}\{\Lambda (\Phi^\dagger\Phi)^2 +
        \lambda (\chi^\dagger\chi)^2 \}  \\
            & + & \lambda'(\Phi^\dagger\Phi)(\chi^\dagger\chi) +
            \lambda''(\Phi^\dagger\chi)(\chi^\dagger\Phi) \\
            & + & \eta'(\Phi^\dagger\Phi)(H^\dagger H) +
            \eta''(\Phi^\dagger H)(H^\dagger\Phi) \\
            & + & \kappa(\chi^\dagger H)(H^\dagger\chi),
    \label{potentialNH}
    \end{array}\end{eqnarray}
where
    \begin{eqnarray}
    V_H = m_H^2H^\dagger H + \frac{1}{2}\eta(H^\dagger H)^2,
    \end{eqnarray}
and
    \begin{eqnarray}
    \Lambda (\Phi^\dagger\Phi)^2 = \lambda_a(\Phi^\dagger\Phi)_1^2 + \lambda_b(\Phi^\dagger\Phi)_{1'}^2 + \lambda_c(\Phi^\dagger\Phi)_2^2,
    \end{eqnarray}
since the product $\Phi^\dagger\Phi$ can be any of the following representations, $(\mathbf{1},1), ~(\mathbf{1'},1)$ or $(\mathbf{2},1)$ of $\mathbb{S}_4\otimes\mathbb{Z}_2$. According to the product rules in Eqs. (\ref{prod1}) - (\ref{prod3}), $(\Phi^\dagger\Phi)_1=|\phi_1|^2+|\phi_2|^2, ~(\Phi^\dagger\Phi)_{1'}=\phi_1^*\phi_2-\phi_2^*\phi_1$, and $(\Phi^\dagger\Phi)_2=(|\phi_2|^2-|\phi_1|^2 ~ ~ \phi_1^*\phi_2+\phi_2^*\phi_1)^T$. The potential in Eq. (\ref{potentialNH}) is symmetric under the interchange of $\phi_1$ and $\phi_2$. Therefore, their vacuum expectation values (vevs) are the same, $\langle\phi_1\rangle = \langle\phi_2\rangle \equiv w$. The vevs $\langle\chi\rangle$ and $\langle H\rangle$, which are denoted as $u$ and $v$, respectively, will be defined to minimize the potential in Eq. (\ref{potentialNH}) in Appendix.A.

After spontaneous symmetry breaking, the Dirac mass matrix of the charged leptons from the Yukawa couplings, $f_{ij}\overline{E}_R^iHL^j$, becomes
    \begin{eqnarray}
        M_{l^-} \sim \left(
        \begin{array}{ccc}
        f_{11}v & 0 & 0 \\
        0 & f_{22}v & 0 \\
        0 & 0 & f_{33}v
        \end{array} \right).\label{smlepton}
    \end{eqnarray}
Thus, this flavor model generates a basis where the matrix of charged lepton masses is diagonal. All the SM particles are $\mathbb{Z}_2$-even as shown in Eq.(\ref{charge_lepton}), so that their Yukawa couplings are protected from the Yukawa couplings with a $\mathbb{Z}_2$-odd Higgs scalar, $\Phi$ or $\chi$.

\section{Minimal Seesaw Mechanism}

\textbf{Minimal Model for Normal Hierarchy}:~
In a minimal model where only two SM singlet righthanded neutrinos $N_1$ and $N_2$ are added to the SM fermions, one zero mass eigenvalue is assigned to three active neutrinos. If the charges of right-handed neutrinos $N_1$ and $N_2$ are given as $(\mathbf{r}_1,q_1)_F=(\mathbf{1'},-1)$ and $(\mathbf{r}_2,q_2)_F=(\mathbf{1},-1)$, then the $\mathbb{S}_4\otimes\mathbb{Z}_2$ invariant Yukawa couplings give rise to the matrix form:
    \begin{eqnarray}
        g_{ij}\Sigma \sim \left(
        \begin{array}{cc}
        0 & g_{12}\chi  \\
        g_{21}\phi_2 & g_{22}\phi_1 \\
        -g_{31}\phi_1 & g_{32}\phi_2
        \end{array} \right).\label{md_nh}
    \end{eqnarray}
The Yukawa coupling $\overline{N}_1(\phi_1~\phi_2)(L_\mu~L_\tau)^T$ is obtained via the tensor product $\mathbf{1'}\times\mathbf{2}\times\mathbf{2}=\mathbf{1'}\times\mathbf{1'}$, whereas the coupling $\overline{N}_2(\phi_1~\phi_2)(L_\mu~L_\tau)^T$ is obtained via the tensor product $\mathbf{1}\times\mathbf{2}\times\mathbf{2}=\mathbf{1}\times\mathbf{1}$. Thus, the different product rules of the same couplings can cause different sizes in the coupling constants, between a pair of $g_{21},~g_{31}$ and a pair of $g_{22},~g_{32}$. This model assumes that $\mathcal{O}(g_{21},~g_{31})<\mathcal{O}(g_{22},~g_{32})$.

When $\phi_1, \phi_2$ and $\chi$ obtain their vacuum expectation values to minimize the potential $V$ in Eq.(\ref{potentialNH}), breaking the symmetry $\mathbb{S}_4\otimes\mathbb{Z}_2$, the process to acquire the vevs $w=\langle\phi_1\rangle=\langle\phi_2\rangle$ and $u=\langle\chi\rangle$ are described in Eq.(\ref{vev_nh}). With the flavor charges of $N_1$ and $N_2$, the mass matrix of the right-handed neutrinos is diagonal, so that $M_{kl}\overline{N}_R^kN_R^l=M_1\overline{N}_1N_1 ~+ ~M_2\overline{N}_2N_2$. According to the seesaw mechanism $ M_\nu = - m_D M_R^{-1} m_D^T$, the mass matrix of light neutrinos $M_\nu$ is
    \begin{eqnarray}
        \left(
        \begin{array}{ccc}
        \frac{g_{12}^2u^2}{M_2} & \frac{g_{12}g_{22}uw}{M_2} & \frac{g_{12}g_{32}uw}{M_2} \\
        \frac{g_{12}g_{22}uw}{M_2} & \frac{g_{21}^2w^2}{M_1}+\frac{g_{22}^2w^2}{M_2} &   -\frac{g_{21}g_{31}w^2}{M_1}+\frac{g_{22}g_{32}w^2}{M_2} \\
        \frac{g_{12}g_{32}uw}{M_2} & -\frac{g_{21}g_{31}w^2}{M_1}+
        \frac{g_{22}g_{32}w^2}{M_2} &
        \frac{g_{31}^2w^2}{M_1}+
        \frac{g_{32}^2w^2}{M_2}
        \end{array} \right),
    \label{mass_nh}
    \end{eqnarray}
where the overall minus sign has been removed by a phase transformation of neutrino fields.

If the masses of light neutrinos have a normal hierarchy, then $m_1=0$. If $u~=~w$, and the Yukawa couplings $g_{ij}$'s are constrained such that $g_{21}=g_{31}$ and $g_{12}=g_{22}=g_{32}$, the mixing matrix of $M_\nu$ is exactly TBM. Besides, there are a priori conditions for TBM mixing that  should be considered. In Eq.(\ref{md_nh}), the zero element is a key to $\theta_{13}=0$ in PMNS, and the opposite sign between the 22 element and 32 element is necessary for $\theta_{23}=\pi/4$. The structure of the Dirac matrix with $m_{12}=0$, $m_{22}m_{32}<0$, and $\langle\phi_1\rangle=\langle\phi_2\rangle$ is derived from $\mathbb{S}_4$ symmetry. In other words, $\theta_{13}=0$ and $\theta_{23}=\pi/4$ are the consequences of $\mathbb{S}_4$ symmetry. However, it is unlikely for a Dirac matrix to have exact zero and exact equalities as a coincidence. It will be shown that slight discrepancies between those elements cause physically significant deviation from the TBM mixing matrix. The non-zero eigenvalues of the light neutrino mass matrix in Eq.(\ref{mass_nh}) are given by $m_2=f(g)(u^2+2w^2)/M_2$ and $m_3=f'(g)~2w^2/M_1$, where $f(g)$ and $f'(g)$ are the factors determined in terms of Yukawa coupling constants $g$'s.

\textbf{Minimal Model for Inverted Hierarchy}:~
When the charges of righthanded neutrinos $N_1$ and $N_2$ are given as $(\mathbf{r}_1,q_1)_F=(\mathbf{1},1)$ and $(\mathbf{r}_2,q_2)_F=(\mathbf{1},-1)$, then the $\mathbb{S}_4\otimes\mathbb{Z}_2$ invariant Yukawa couplings give rise to
    \begin{eqnarray}
        g_{ij}\Sigma \sim \left(
        \begin{array}{cc}
        g_{11}H & g_{12}\chi \\
        0 & g_{22}\phi_1 \\
        0 & g_{32}\phi_2
        \end{array} \right).\label{md_ih}
    \end{eqnarray}
With such charge assignment, the mass matrix of righthanded neutrinos is  diagonal; $M_{kl} \overline{N}_R^k N_R^l=M_1\overline{N}_1N_1+M_2\overline{N}_2N_2$.
According to the seesaw mechanism $ M_\nu = - m_D M_R^{-1} m_D^T$, the mass matrix of light neutrinos $M_\nu$ is
    \begin{eqnarray}
    \left(
        \begin{array}{ccc}
        \frac{g_{11}^2v^2}{M_1}+
        \frac{g_{12}^2u^2}{M_2} & \frac{g_{12}g_{22}uw}{M_2} & \frac{g_{12}g_{32}uw}{M_2} \\
        \frac{g_{12}g_{22}uw}{M_2} & \frac{g_{22}^2w^2}{M_2} & \frac{g_{22}g_{32}w^2}{M_2} \\
        \frac{g_{12}g_{32}uw}{M_2} & \frac{g_{22}g_{32}w^2}{M_2} &
        \frac{g_{32}^2w^2}{M_2}\label{mass_ih}
        \end{array} \right),
    \end{eqnarray}
where $u=\langle\chi\rangle, v=\langle H\rangle$, and $w=\langle\phi_1\rangle=\langle\phi_2\rangle$. When taking the seesaw mechanism, the two zero elements in Eq.(\ref{md_ih}) are necessary for $\theta_{13}=0$, and $g_{22}=g_{32}$ is necessary for $\theta_{23}=\pi /4$. In other words, the $\mathbb{S}_4\otimes\mathbb{Z}_2$ flavor symmetry results in the texture zeros and the equalities, which are the necessary conditions for TBM mixing in PMNS. Non-zero eigenvalues of light neutrino mass matrix in Eq.(\ref{mass_ih}), $m_1$ and $m_2$, are given by
    \begin{eqnarray}
    && m_2+m_1 = f(g) \left( \frac{v^2}{M_1} + \frac{u^2+2w^2}{M_2} \right) \label{eigen_ih} \\
    && m_2-m_1 = f'(g) \left( \left( \frac{v^2}{M_1} + \frac{u^2+2w^2}{M_2} \right)^2 - \frac{8u^2w^2}{M_1M_2} \right)^{1/2},
    \nonumber
    \end{eqnarray}
where $f(g)$ and $f'(g)$ can be determined in terms of Yukawa coupling constants $g$'s.
In addition, possible deviations from those specific elements in Dirac mass matrix can cause a phenomenological deviation from TBM mixing in the PMNS matrix.

\section{Symmetry breaking and deviation from Tri-Bi Maximal mixing}

It is necessary to examine the possible aspects that can appear when the zeros and equalities generated by the flavor symmetry are collapsed during symmetry breaking. Therefore, some simple cases will be discussed. In a trial model, it can be assumed to have a right-handed mass matrix as follows:
    \begin{eqnarray}
        M_R = \left(
        \begin{array}{cc}
        1 & 0  \\
        0 & 10^2
        \end{array} \right)M_1,\label{dev_mr}
    \end{eqnarray}
and $f(g)$ and $f'(g)$, in normal hierarch case or in inverted hierarch case Eq.(\ref{eigen_ih}), are assumed to be one, for simplicity.
It is unlikely to have an exact equality $u=w$, or to have an exact zero element, as shown in Eq.(\ref{md_nh}) and Eq.(\ref{md_ih}), after a series of spontaneous symmetry breaking mechanisms. In order to examine the effects of symmetry breaking to a deviation from low energy TBM mixing matrix, the Dirac mass matrix for normal hierarchy is assumed to have the following form:
    \begin{eqnarray}
        m_D = \left(
        \begin{array}{cc}
        t & u  \\
        \gamma w & w \\
        -\gamma w & w
        \end{array} \right),\label{dev_md_nh}
    \end{eqnarray}
where $t<u$ or $w$, and $t$ is a small value that can occur from  $\mathbb{S}_4\otimes\mathbb{Z}_2$-violating Yukawa coupling. The factor $\gamma$ is a Yukawa coupling ratio that affects the mass ratio at low energy, which can arise from a $\mathbf{1'}\times\mathbf{1'}$ product weaker than $\mathbf{1}\times\mathbf{1}$. The curve (a) in Fig.1 describes the change in $|U_{e2}|$ as the ratio $u/w$ varies from 0 to 1.5 for $t=0$, even though the change in $|U_{e2}|$ turns out to be independent of $t$. On the other hand, the variation in $u/w$ does not cause any change in $|U_{e3}|$ or $|U_{\mu 3}|$. The curves (b) and (c) in Fig.1 describe the changes in $|U_{e3}|$ and the change in $|U_{\mu 3}|$, respectively, as the ratio $t/w$ varies from 0 to 1, while keeping $u/w=1$. Under the given conditions, a range of $u/w$, $0.82\leq u/w \leq 1.07$, can lead to $|U_{e2}|$ at the 90\% CL. The range of $|U_{e3}|$ at the 90\% CL matches with $t/w \leq 0.35$, while the range of $|U_{\mu3}|$ at 90\% CL matches with $t/w \leq 0.46$. The size of $\gamma$ does not affect the mixing angles in PMNS, but does affect the mass ratios.
    \begin{figure*}
        \resizebox{150mm}{!}
        {\includegraphics[width=0.75\textwidth]{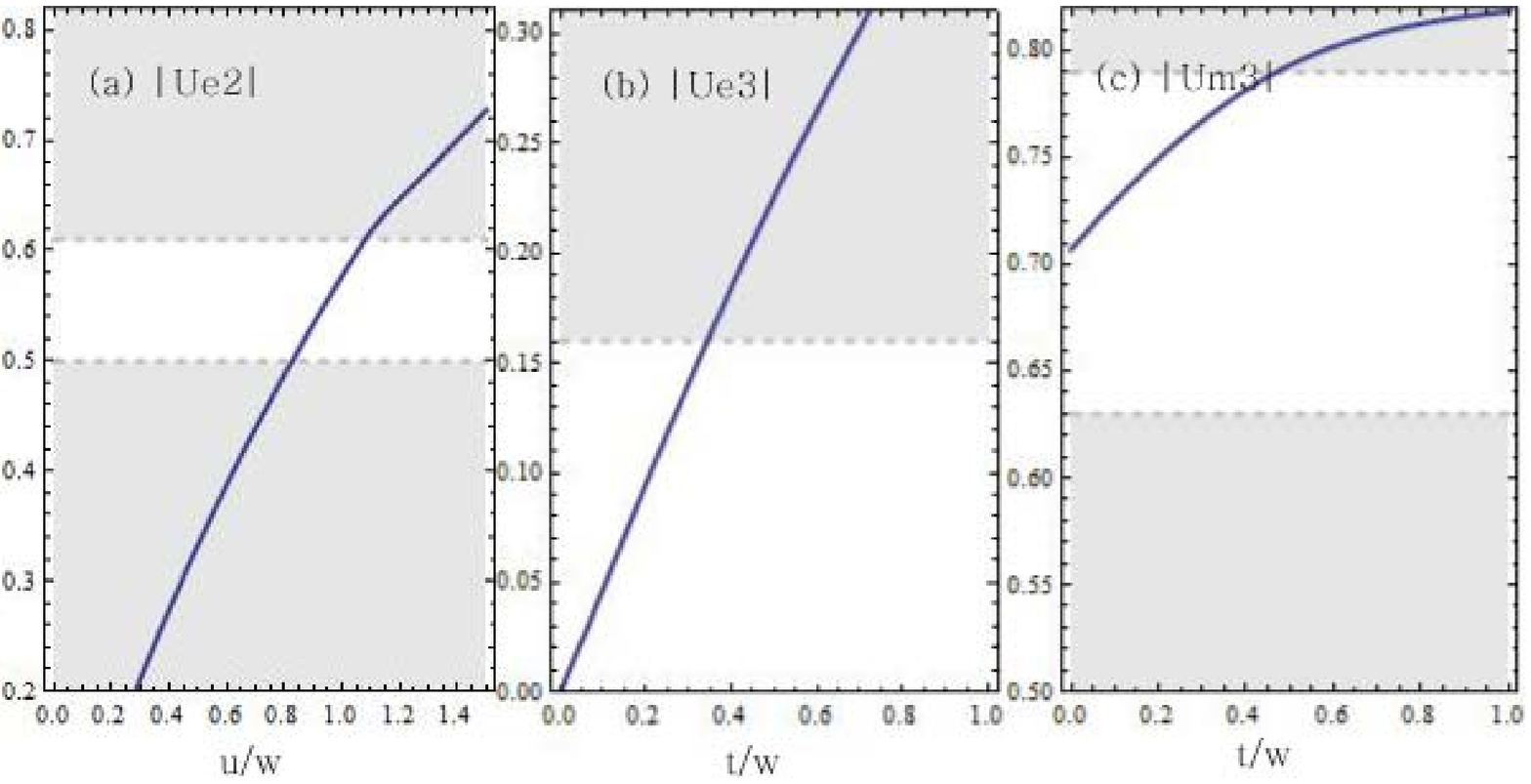}}
        \caption{\label{fig:nh3}
        Unshaded regions exhibit the allowed ranges in the elements of PMNS matrix at the 90\% CL. The curve (a) $|U_{e2}|$ is obtained at $t=0$ in Eq.(\ref{dev_md_nh}), and its intersections with the bounds are at $u/w=0.82$ and at $u/w=1.07$. The curve (b) $|U_{e3}|$ is obtained at $u/w=1$, and its intersection with the bound is at $t/w=0.35$, whereas the curve (c) $|U_{\mu3}|$ has an intersection with the bound at $t/w=0.46$.  }
    \end{figure*}

For inverted hierarchy, the Dirac mass matrix can be supposed to have the following form:
    \begin{eqnarray}
        m_D = \left(
        \begin{array}{cc}
        v & u  \\
        g_1 t & w \\
        g_2 t & w
        \end{array} \right),\label{dev_md_ih}
    \end{eqnarray}
where $t < u,$ or $w$. Yukawa couplings $g_1$ and $g_2$ are assumed to be very small so that $g_1 t$ and $g_2 t$ are much smaller than $v$. The curve (a) in Fig.2 describes the change in $|U_{e2}|$ as the ratio $v/w$ varies from 0 to 1.5 for $u=w$ and $t=0$. Under the given conditions, a range of $v/w$, $0.135\leq v/w \leq 0.165$, can lead to $|U_{e2}|$ at the 90\% CL. Since $v$ is the vacuum expectation value of the Higgs scalar in the Standard Model, this model does not cause any new physics below approximately 1 TeV.
The curves (b) and (c) in Fig.1 describe the changes in $|U_{e3}|$ and  $|U_{\mu 3}|$, respectively, when the 2-1 and 3-1 elements obtain a non-zero value due to $\mathbb{S}_4\otimes\mathbb{Z}_2$-violating coupling, and their relative ratio $g_1/g_2$ varies from 0 to 10. The range of $|U_{e3}|$ at the 90\% CL matches with $g_1/g_2 \leq 3.8$, whereas the range of $|U_{\mu 3}|$ at the 90\% CL matches with $g_1/g_2 \leq 4.7$.

    \begin{figure*}
        \resizebox{150mm}{!}
        {\includegraphics[width=0.75\textwidth]{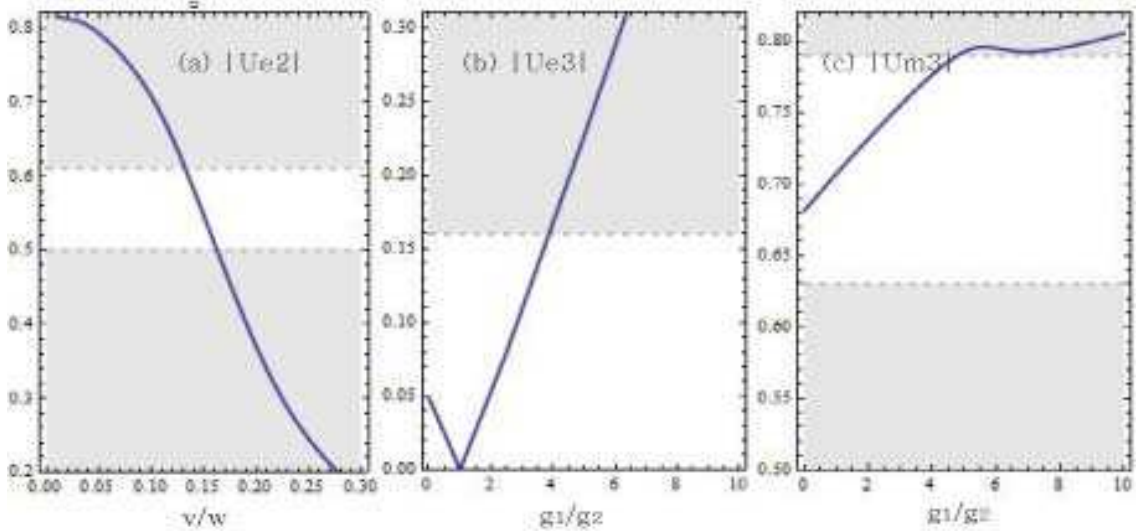}}
        \caption{\label{fig:ih3}
        Unshaded regions exhibit the allowed ranges in the elements of the PMNS matrix at the 90\% CL. The curve (a) $|U_{e2}|$ is obtained at $t=0$ in Eq.(\ref{dev_md_ih}), and its intersections with the bounds are at $v/w=0.135$ and at $v/w=0.165$. The curve (b) $|U_{e3}|$ is obtained at $v/w=0.15$, and its intersection with the bound is at $g_1/g_2=3.8$, whereas the curve (c) $|U_{\mu3}|$ has an intersection with the bound at $g_1/g_2=4.6$.}
    \end{figure*}

\section{Conclusion}

A model of lepton masses was constructed in the framework of a discrete flavor symmetry $\mathbb{S}_4\otimes\mathbb{Z}_2$. Each family of SM leptons or each of the right-handed neutrinos, is distinguished from each other by their flavor charges. However, a pair of $L_\mu$ and $L_\tau$ are assigned to the representation $\mathbf{2}$ of $\mathbb{S}_4$ so that they cannot be distinguished under the Yukawa interaction, and so are a pair of $\mu_R$ and $\tau_R$, indicating $\mu-\tau$ symmetry. When the symmetry is accompanied with $\mathbb{Z}_2$ symmetry, we can obtain the basis where the mass matrix of charged leptons is diagonal, as shown in Eq.(\ref{smlepton}). Since the SM particles are all $\mathbb{Z}_2$-even and the additional scalar particles are $\mathbb{Z}_2$-odd, there are no extra Yukawa couplings of charged leptons found. The model contains only two righthanded neutrinos so that the couplings with three scalar fields and three lefthanded neutrinos generate a $2\times3$ Yukawa matrix. This model can be characterized by a few noteworthy points. First, the group theoretical properties of $\mathbb{S}_4$ gives rise to $\mu-\tau$ symmetry, accordingly TBM mixing, and the diagonal mass matrix of charged leptons. Second, the type of neutrino mass spectrum, normal or inverted hierarchy, is resulted in by the flavor symmetry. The flavor charges of righthanded neutrinos play a key role in determining the order of light neutrino masses. Third, it is possible to build the mass matrices with a significantly small content of scalar particles under a non-abelian discrete flavor symmetry $\mathbb{S}_4\otimes\mathbb{Z}_2$.

The texture zeros or the equalities due to the symmetry in a Yukawa matrix appear to be barely protected, once the symmetry undergoes a chain of breakdown. It is natural for PMNS to obtain a deviation from TBM mixing when the symmetry of the Yukawa matrix is broken. Fig.\ref{fig:nh3} and Fig.\ref{fig:ih3} show a schematic diagram of some aspects that can appear in the correlation between the order of symmetry breaking and the deviation from TBM.

\begin{acknowledgments}
KS wishes to thank the members in the Physics Department at Chonnam National
University for their warm hospitality.

\end{acknowledgments}

\appendix

\section{Higgs Potential}

The $\mathbb{S}_4\otimes\mathbb{Z}_2$ invariant Higgs potential in Eq.(\ref{potentialNH}) can be rephrased in terms of $\{\phi_i, \phi_i^\dagger\}$ with $i=1$ and 2, $\chi$, and $H$.
    \begin{eqnarray}
    &&V~(~\phi_i,\phi_i^\dagger,\chi,H~) \nonumber \\
    &&= m_H^2|H|^2 + \frac{1}{2}\eta|H|^4
    + m_\varphi^2 (|\phi_1|^2+|\phi_2|^2) + m_\chi^2|\chi|^2  \nonumber\\
    &&+ \frac{1}{2}(\lambda_a+\lambda_c)(|\phi_1|^4+|\phi_2|^4)
    + (\lambda_a+\lambda_b)|\phi_1|^2|\phi_2|^2 \nonumber \\
    &&+ \frac{1}{2}(\lambda_c-\lambda_b)(\phi_1^{*2}\phi_2^2 + \phi_2^{*2}\phi_1^2) \\
    &&+ \frac{1}{2}\lambda|\chi|^4+ (\lambda'+\lambda'')(|\phi_1|^2+|\phi_2|^2)|\chi|^2 \nonumber \\
    &&+ (\eta'+\eta'')(|\phi_1|^2+|\phi_2|^2)|H|^2 +\kappa|\chi|^2|H|^2. \nonumber
    \end{eqnarray}
When the Higgs particles obtain their vacuum expectation values such that $\langle\chi\rangle=u$, $\langle H\rangle=v$, and $\langle\phi_1\rangle=\langle\phi_2\rangle=w$ where $u, v,$ and $w$ are real, the minimal potential can be expressed as follow:
    \begin{eqnarray}
    && V_{min}~(u,v,w) ~= ~m_H^2v^2 + m_\chi^2u^2 + 2m_\varphi^2w^2   \nonumber\\
    && + \frac{1}{2}\lambda u^4 + \frac{1}{2}\eta v^4+ ~2(\lambda_a+\lambda_c)w^4 \\
    && + 2(\lambda'+\lambda'')u^2w^2 +2(\eta'+\eta'')v^2w^2 +
    \kappa u^2v^2, \nonumber
    \end{eqnarray}
which satisfies
    \begin{eqnarray}
    &&\frac{\partial\mathcal{L}}{\partial v} =
    2v\{m_H^2 +\kappa u^2 +\eta v^2 +\Lambda_c w^2\}=0 \nonumber \\
    &&\frac{\partial\mathcal{L}}{\partial u} =
    2u\{m_\chi^2 +\lambda u^2 +\kappa v^2 +\Lambda_b w^2\}=0 \\
    &&\frac{\partial\mathcal{L}}{\partial w} =
    2w\{2m_\varphi^2 +2\Lambda_b u^2 +\Lambda_c v^2 +\Lambda_a w^2\}=0, \label{vev_nh}
    \nonumber
    \end{eqnarray}
where $\Lambda_a=2(\lambda_a+\lambda_b)$, $\Lambda_b=2(\lambda'+\lambda'')$ and $\Lambda_c=2(\eta'+\eta'')$.
Therefore, the vacuum expectation values $u, v$ and $w$ are
\begin{widetext}
    \begin{eqnarray}
    && u^2=\{(\Lambda_b\Lambda_c-\kappa\Lambda_a)m_H^2
        + (\eta\Lambda_a-\lambda_c^2)m_\chi^2
        + (2\kappa\Lambda_c-2\eta\Lambda_b)m_\varphi^2 \}/D
        \nonumber \\
    && v^2=\{(\lambda\Lambda_a-2\Lambda_b^2)m_H^2
        + (2\Lambda_b\Lambda_c-\kappa\Lambda_a)m_\chi^2
        + (2\kappa\Lambda_b-2\lambda\Lambda_c)m_\varphi^2 \}/D
        \label{vev_ih} \\
    && w^2=\{(2\kappa\Lambda_b-\lambda\Lambda_c)m_H^2
        + (\kappa\Lambda_c-2\eta\Lambda_b)m_\chi^2
        + (2\eta\lambda-2\kappa^2)m_\varphi^2 \}/D
        \nonumber \\
    && D=(\kappa^2-\eta\lambda)\Lambda_a+2\eta\Lambda_b^2
        - 3\kappa\Lambda_b\Lambda_c+\lambda\Lambda_c^2. \nonumber
    \end{eqnarray}
\end{widetext}

\end{document}